\documentclass[aps,preprint,preprintnumbers,amsmath,amssymb]{revtex4}

\usepackage{graphicx}
\usepackage{dcolumn}
\usepackage{bm}
\usepackage{color}
\usepackage{natbib}

\newcommand*{\cm}[1]{#1~cm$^{-1}$}

\begin{document}


\title{Band Structure of Two-dimensional Dirac Semimetal from Cyclotron Resonance}

\author{A.~M.~Shuvaev$^{1}$}
\author{V.~Dziom$^{1}$}
\author{N.~N.~Mikhailov$^{2,3}$}
\author{Z.~D.~Kvon$^{2,3}$}
\author{Y.~Shao$^{4}$}
\author{D.~N.~Basov$^{4}$}
\author{A.~Pimenov$^{1}$}

\affiliation{$^{1}$Institute of Solid State Physics, Vienna University of
Technology, 1040 Vienna, Austria \\
$^{2}$Rzhanov Institute of Semiconductor Physics, 630090 Novosibirsk, Russia \\
$^{3}$Novosibirsk State University, Novosibirsk 630090, Russia \\
$^{4}$Department of Physics, Columbia University, New York 10027, USA}

\begin{abstract}
Knowing the band structure of materials is one of the prerequisites to understand their properties. Therefore, especially in the last decades, angle-resolved photoemission spectroscopy (ARPES) has become a highly demanded experimental tool to investigate the band structure. However, especially in thin film materials with a layered structure and several capping layers, access to the electronic structure by ARPES is limited. Therefore, several alternative methods to obtain the required information have been suggested. Here, we directly invert the results by cyclotron resonance experiments to obtain the band structure of a two-dimensional (2D) material. This procedure is applied to the mercury telluride quantum well with critical thickness which is characterized by a 2D electron gas with linear dispersion relations. The Dirac-like band structure in this material could be mapped both on the electron and on the hole side of the band diagram. In this material, purely linear dispersion of the hole-like carriers is in contrast to detectable quadratic corrections for the electrons.
\end{abstract}

\date{\today}


\maketitle

\section{Introduction}
An electronic band structures provides the relationship between quasiparticle energy and momentum and it can be understood as the material fingerprint in the reciprocal space. Filled, half-filled or empty bands provide immediate information about the metallic or insulating properties of a material in question. A standard modern method to obtain the band structure of solids is provided by ARPES~\cite{damascelli_rmp_2003}. In ARPES experiments, electrons escape from the solid after absorbing a photon of well-defined energy. Kinetic energy and momentum of the electrons are analyzed, which results in the band structure of the material. As the electron mean free path in solids are typically in the {\AA}ngstr\"{o}m range, basically only the sample area close to the surface can be investigated.

The surface of thin film materials is often covered by several additional layers. These are required technologically, for example, as a buffer-, doping-, or capping-layers. These layers absorb the photo-emitted electrons, thereby making the investigation of the band structure by ARPES difficult. To overcome this problem, several alternative methods have been suggested, like the analysis of the cyclotron mass~\cite{novoselov_nature_2005, zhang_nat_2005, zhang_nphys_2011, minkov_prb_2014} or density of states via capacitance experiments~\cite{kozlov_prl_2016, kozlov_jetpl_2016}. In such cases, the experimentalists compare the theoretical calculations~\cite{chu_book, yu_book} with experimental parameters obtained in the experiment~\cite{hancock_prl_2011, orlita_np_2014,zoth_prb_2014, dantscher_prb_2015, akrap_prl_2016}.  Especially for metals, the de Haas-van Alphen effect has been proven to provide valuable information about the Fermi surfaces \cite{ashcroft_book, shoenberg_book, minkov_prb_2013, minkov_prb_2014, minkov_prb_2015, minkov_prb_2016}. Recently, new access to the band structure in solids have been suggested~\cite{vampa_prl_2015}, which utilizes nonlinear optical effects. The momentum-dependent band gap is obtained by analyzing the recombination of the electron-hole pairs generated by a femtosecond laser pulse.

In this work, we directly invert the data by cyclotron resonance experiments to obtain the band structure of a two-dimensional (2D) Dirac semi-metal. The quantitative analysis is performed in 2D HgTe quantum wells combined with the variation of charge density by the gate voltage.

The access to the dispersion relations $E(k)$ via the cyclotron experiments is based on the expression connecting the cyclotron resonance $\Omega_c$ and the details of the band structure at the Fermi energy $E_F$~\cite{ashcroft_book, neto_rmp_2009}
\begin{equation}\label{eqCR}
  \Omega_c^{-1}= \frac{m_c}{eB}=  \frac{\hbar^2}{2 \pi eB}\frac{\partial A}{\partial E}  \Big|_{E=E_F} \quad.
\end{equation}
Here $B$ is the magnetic field, $A$ is the area in the reciprocal space enclosed by the contour of the constant energy $E$, and $m_c=(\hbar^2/2\pi)\partial A/\partial E$ is the cyclotron mass. (See, e.g. Ref.~\cite{ariel_arxiv_2012} for comparison of definitions of the effective mass.) In a 2D case and within a reasonable approximation, the dispersion relations $E(k)$ may be assumed as rotation-symmetric, which gives $A=\pi k^2$. Equation (\ref{eqCR}) thus can be rewritten as:
\begin{equation}\label{eqCR2}
  \Omega_c^{-1}= \frac{\hbar k_F}{eB} \frac{1}{v_F (k_F)} =  \frac{\hbar^2 k_F}{eB}\frac{\partial k}{\partial E} \Big|_{E=E_F} \quad,
\end{equation}
which represents the 2D band structure in the differential form. Here $ v_F = ({\partial E}/{\partial k})_{E=E_F}/\hbar  $ is the Fermi-velocity. In the same approximation, the quasi-particle wave vector at the Fermi energy $k_F$ may be calculated from the charge density $n_{2D}$ as  $k_F = \sqrt{2\pi n_{2D}}$. We note here that within the same approximations neither $E_F$ nor $k_F$ are magnetic field dependent. Therefore, according to Eq.~(\ref{eqCR2}) in the quasi-classical approximation the cyclotron frequency depends linearly on magnetic field irrespective of the specific form of the dispersion relations. For the present samples this dependence is demonstrated for $B < 1$~T in Fig.~\ref{figir} below.

The remaining experimental problem is to change the position of the Fermi energy or, equivalently, of the 2D density in a broad range. This can be achieved using a transparent metallic gate on the top of the sample~\cite{shuvaev_apl_2013, shuvaev_prl_2016}. An alternative possibility which can also be applied to mercury telluride is to change the charge density via illumination by visible light \cite{ikonnikov_sst_2011, kvon_jetpl_2012, olbrich_prb_2013, zoth_prb_2014, dziom_2d_2017}.

Mercury telluride (HgTe) films with critical thickness $d \approx 6.3-6.6$~nm
are characterized by a 2D band structure with a Dirac cone in the vicinity of the Fermi energy~\cite{bernevig_science_2006, konig_science_2007, buttner_nphys_2011}. Compared to the closely similar case of graphene~\cite{neto_rmp_2009}, in which the Dirac cones are fourfold degenerate due to spin and valley degeneracy, in HgTe, the cone is only twofold degenerate. Another important fact is that, contrary to graphene, the Dirac cone in the 2D HgTe is predicted to be asymmetric with respect to electron and hole sides~\cite{qi_prb_2008}. This is one more argument why information about the actual band structure is important.

\begin{figure}[tbp]
\includegraphics[width=0.8\linewidth]{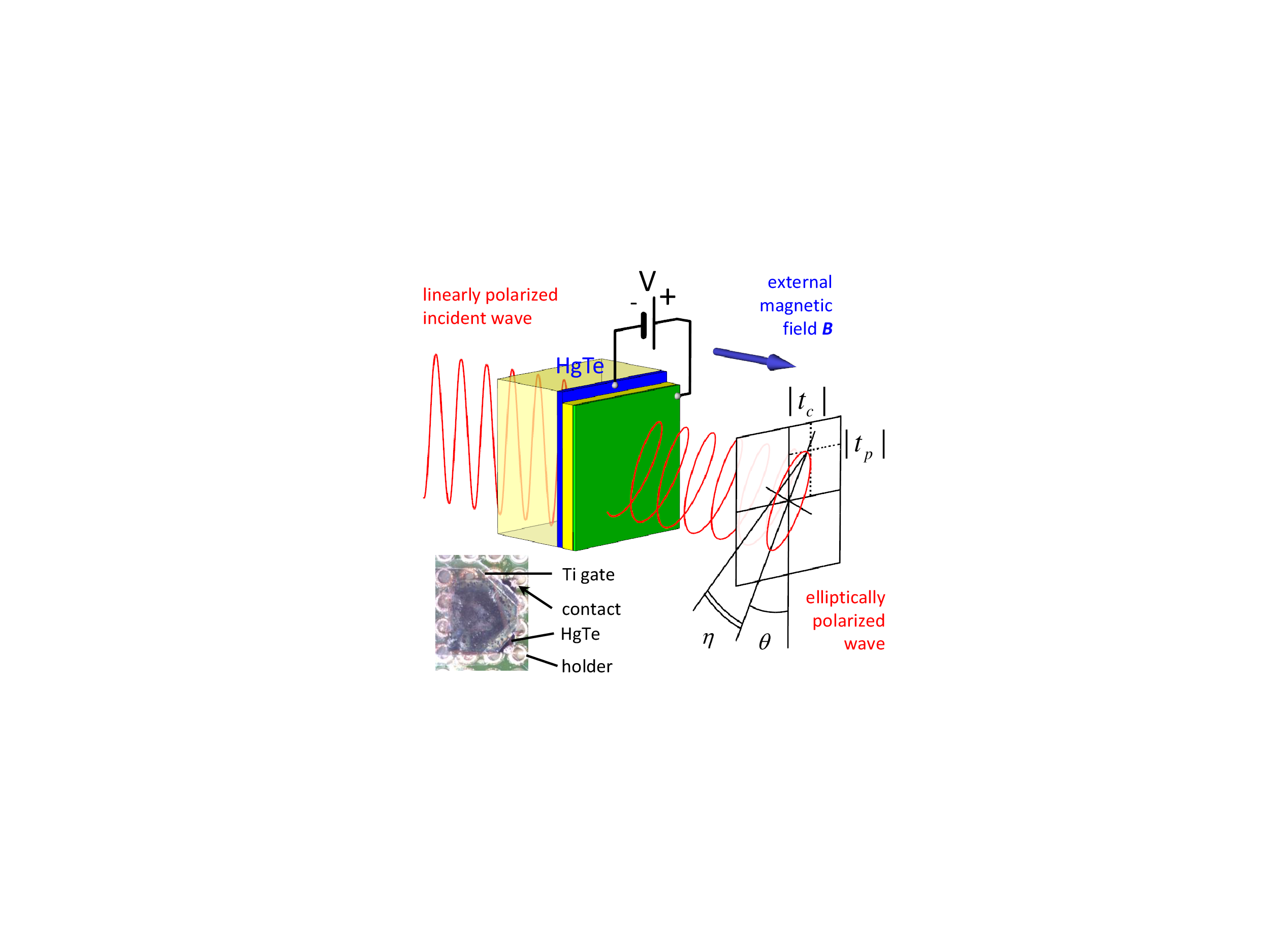}
\caption{\textbf{Schematic drawing of the magneto-optical terahertz transmission experiment.} $\theta$ is the Faraday rotation angle, $\eta$ is the ellipticity, $t_p$ and $t_c$ are the transmission amplitudes in the parallel and crossed polarizers, respectively. The inset shows the photograph of the sample with \textit{ex-situ} gate. } \label{figexp}
\end{figure}

To obtain an experimental approach to the band structure using Eq.~(\ref{eqCR2}), the cyclotron mass must be measured with simultaneous control of the 2D density. In the present work, this has been achieved using a continuous-wave terahertz technique with \textit{ex-situ} gates (Fig.~\ref{figexp}).

\section{Experimental details}

\subsection{Experimental technique and data analysis}

\begin{figure}[tbp]
\includegraphics[width=0.65\linewidth]{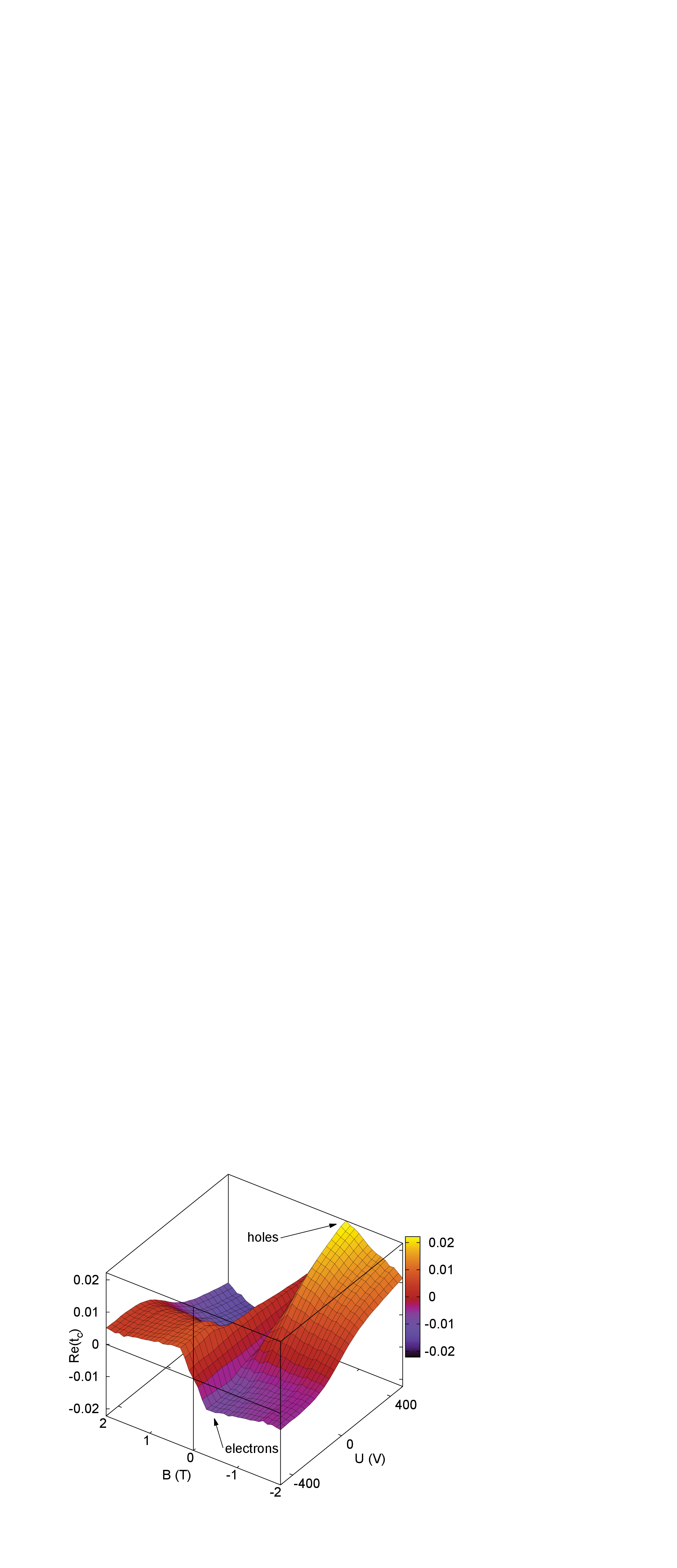}
\caption{\textbf{Cyclotron resonance in the terahertz range.}  Three-dimensional (3D) presentation of the measured transmission at 320 GHz (sample \#2)  in crossed polarizers geometry and in a broad range of magnetic fields and gate voltages. In a simple approximation, the crossed transmission is proportional to the Faraday rotation angle $\theta$. } \label{fig3da}
\end{figure}

Magneto-optical experiments at submillimeter frequencies were performed in a Mach-Zehnder interferometer~\cite{volkov_infrared_1985}, which allows
measurements of the amplitude and the phase shift in a geometry with
controlled polarization of radiation. Continuous
radiation is produced by backward wave oscillators (BWO). The polarization
of the beam is controlled by wire grid polarizers. The sample is placed
in the helium-cooled cryostat with the superconducting split-coil magnet with polypropylene windows.
The main measurement mode in the present work is to fix the frequency
of the generated radiation and to measure both the transmission
amplitude and the phase difference of the transmitted signal as a function
of the applied magnetic field and the gate voltage. In each experiment two measurements are performed: one with the analyzer setting the same as the incident polarization (parallel polarizers geometry, $t_p$) and one with the analyzer rotated by 90$^{\circ}$
(crossed polarizers geometry, $t_c$).

Magneto-optical transmission experiments in the infrared frequency range 30~cm$^{-1}<\nu<700$~cm$^{-1}$ have been performed using a Fourier-transform spectrometer equipped with a superconducting magnet and a He-cooled bolometer.

Figure \ref{fig3da} shows a typical example of the transmission in the crossed polarizers measured in the broad range of the external magnetic fields and the gate voltages. The experimental curves are dominated by a resonance-like feature around $\pm 0.4$~T, corresponding to a cyclotron resonance.

As the phase shift of the transmitted signal is measured in the experiment, the sign of the charge carriers can be obtained. This may be seen from the following arguments. The type of the charge carriers determines the sign of the cyclotron resonance via $\Omega_c= {eB}/{m_c}$. As follows from Eqs. (\ref{eqtr}, \ref{drude}), the relative phase $\varphi$ of the transmission in crossed polarizers depends on the sign of $\Omega_c$. In Fig. \ref{fig3da} the real part of $t_c$ is shown which changes sign with the direction of magnetic field and on the transmission from electron to holes ( i.e. positive vs. negative voltages).  Therefore, in this experiment we may directly access both, electron and hole parts of the band structure.

The polarization rotation $\theta$ and the ellipticity $\eta$  can be considered as intermediate experimental properties and they are
obtained from the transmission data as
\begin{eqnarray}
&& \tan(2\theta)=2\Re(\chi)/(1-|\chi|^2)\ , \\
&& \sin(2\eta)=2\Im(\chi)/(1+|\chi|^2)\
\end{eqnarray}
with $\chi=t_c/t_p$.

\begin{figure}[tbp]
\includegraphics[width=0.75\linewidth]{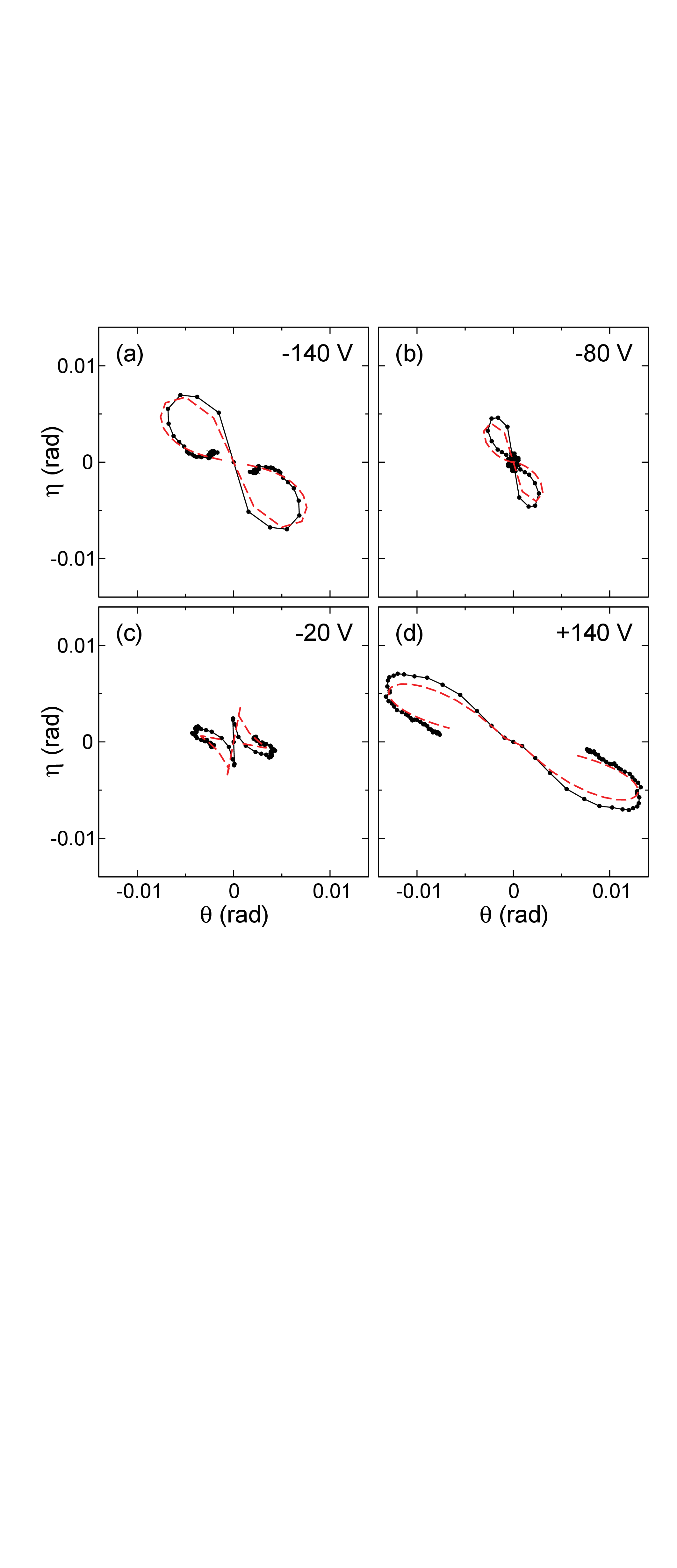}
\caption{\textbf{Complex-plane presentation of the cyclotron resonance.}  Complex presentation of the Faraday rotation, $\theta$, and ellipticity, $\eta$, at selected gate voltages demonstrating the transition from hole-like (a) to electron-like (d) carriers (sample \#2, $\nu= 320$~GHz). Symbols - experimental data, red lines - model fit using Drude conductivity, Eq.~(\ref{drude}), and black lines are to guide the eye.} \label{fig3db}
\end{figure}

To obtain the dynamic conductivity from the measured complex transmission, the procedure similar to Berreman formalism~\cite{berreman_josa_1972} has been utilized. Within this procedure the electromagnetic radiation is represented via the 4D vector of transversal components of the \textit{ac} electric and magnetic fields. In the next steps the complex transmission is calculated via the Maxwell equations and the boundary conditions using the $4 \times 4$ propagation matrices which connect the electromagnetic fields in different points in space~\cite{shuvaev_sst_2012}.

In earlier experiments \cite{shuvaev_prl_2011, shuvaev_sst_2012} a numerical algorithm has been utilized to invert the transmission expressions. Recently, the analytical formula has been developed for the case of a thin film on a substrate \cite{dziom_2d_2017}, thus strongly simplifying the procedure. For example, the final equation for the off-diagonal 2D conductivity may be written as:
\begin{equation}\label{eqtr}
  \sigma_{xy}= \frac{2 \sqrt{\varepsilon} t_{c} e^{-i a \omega/c}} {Z_0  (t_{p}^2+t_{c}^2)(\sqrt{\varepsilon}\cos{\beta}-i\sin{\beta})} \quad,
\end{equation}
where $a$ is the substrate thickness, $\varepsilon$ is the dielectric
permittivity of the substrate, $d$ is the film thickness, $\omega = 2\pi
f$ is the angular frequency, $\beta=
\sqrt{\varepsilon}a{\omega}/{c}$, and $Z_0 \approx 377$~$\Omega$ is the impedance of free space.

\subsection{Sample preparation}

Mercury telluride quantum wells have been grown on (013) oriented GaAs substrates by molecular beam epitaxy~\cite{kvon_ltp_2009}. Three samples with thickness close to critical have been investigated. For the terahertz experiments sample \#1 with $d=6.3$~nm and sample \#2 with $d=6.6$~nm, these have been used in combination with an \textit{ex-situ} gate. The gate on both samples has been prepared using a mylar film with $d=6 \mu$m as an insulating barrier and a semi-transparent metallized Ti-film as a gate ($R=600 \Omega/ \Box$). In the experiment, the gate conductivity is seen as a magnetic field-independent and frequency independent contribution to $\sigma_{xx}$. No measurable effect of gate on the Hall conductivity has been observed, which agrees well with low mobility of the carriers in the gate electrode. For the infrared experiments, sample \#3 with $d=6.6$~nm without gate has been used.

\section{Results and discussion}

\subsection{Cyclotron resonance}

To obtain such parameters of the charge carriers like density, mobility, and effective cyclotron mass, the experimental spectra were fitted using the Drude model for the magnetoelectric conductivity \cite{tse_prb_2011, tkachov_prb_2011, shuvaev_prb_2013}:
\begin{eqnarray}
&& \sigma_{xx} = \sigma_{yy} = \frac{1 - \imath \omega \tau}
{(1 - \imath \omega \tau)^2 + (\Omega_c \tau)^2} \sigma_0 \,, \label{drude}\\
&& \sigma_{xy} = -\sigma_{yx} = \frac{\Omega_c \tau}
{(1 - i \omega \tau)^2 + (\Omega_c \tau)^2} \sigma_0 \,. \nonumber
\end{eqnarray}
Here, $\tau$ is the relaxation time of the charge carriers,
$\sigma_0 = n_{2D} e^2 \tau / m_{c}$ is the 2D conductivity,
and $B$ is the external magnetic field perpendicular to the film surface.
A few examples of such fits are shown in Fig.~\ref{fig3db}(a-d) presenting the Faraday rotation and ellipticity in the complex plane with magnetic field as a parameter. In Fig.~\ref{fig3db}(a) with the gate voltage $U=-140$~V, the spectra are dominated by a single hole contribution and the curve has a characteristic tilted "figure-eight" shape. This shape may be understood in a thin film approximation, Eq.~(\ref{eqtr}), where $\theta + i\eta \approx t_c/t_p \approx t_c \sim \sigma_{xy}$. Further on, for $\omega \tau \gg 1$ Eq.~(\ref{drude}) may be written as $\sigma_{xy} \sim \Omega_c/(\Omega_c^2-\omega^2-2i\omega/\tau)$ which leads to a resonance at $\Omega_c = \omega$. Due to the imaginary term $2i\omega/\tau$ in the denominator the characteristic form in the complex-plane presentation is a figure-eight curve with two lobes corresponding to $B>0$ and $B<0$, respectively (Fig.~\ref{fig3db}). The complex-plane presentation is especially advantageous if contributions from different charge carriers should be visualized.

With increasing gate voltage towards zero, the amplitude of the hole response is gradually suppressed (Fig.~\ref{fig3db}(b)). In addition, close to the point of origin a tiny feature is observed corresponding to an electronic contribution which starts to grow. We attribute a simultaneous appearance of the electron and hole contribution to inhomogeneities in the film and in the insulating layer. For further increasing voltages (Fig.~\ref{fig3db}(c)), the electronic contribution becomes gradually larger than the hole response. Finally, for large positive voltages (Fig.~\ref{fig3db}(d)), only the electron contribution is detected in the cyclotron resonance spectra.

\begin{figure}[tbp]
\includegraphics[width=.90\linewidth]{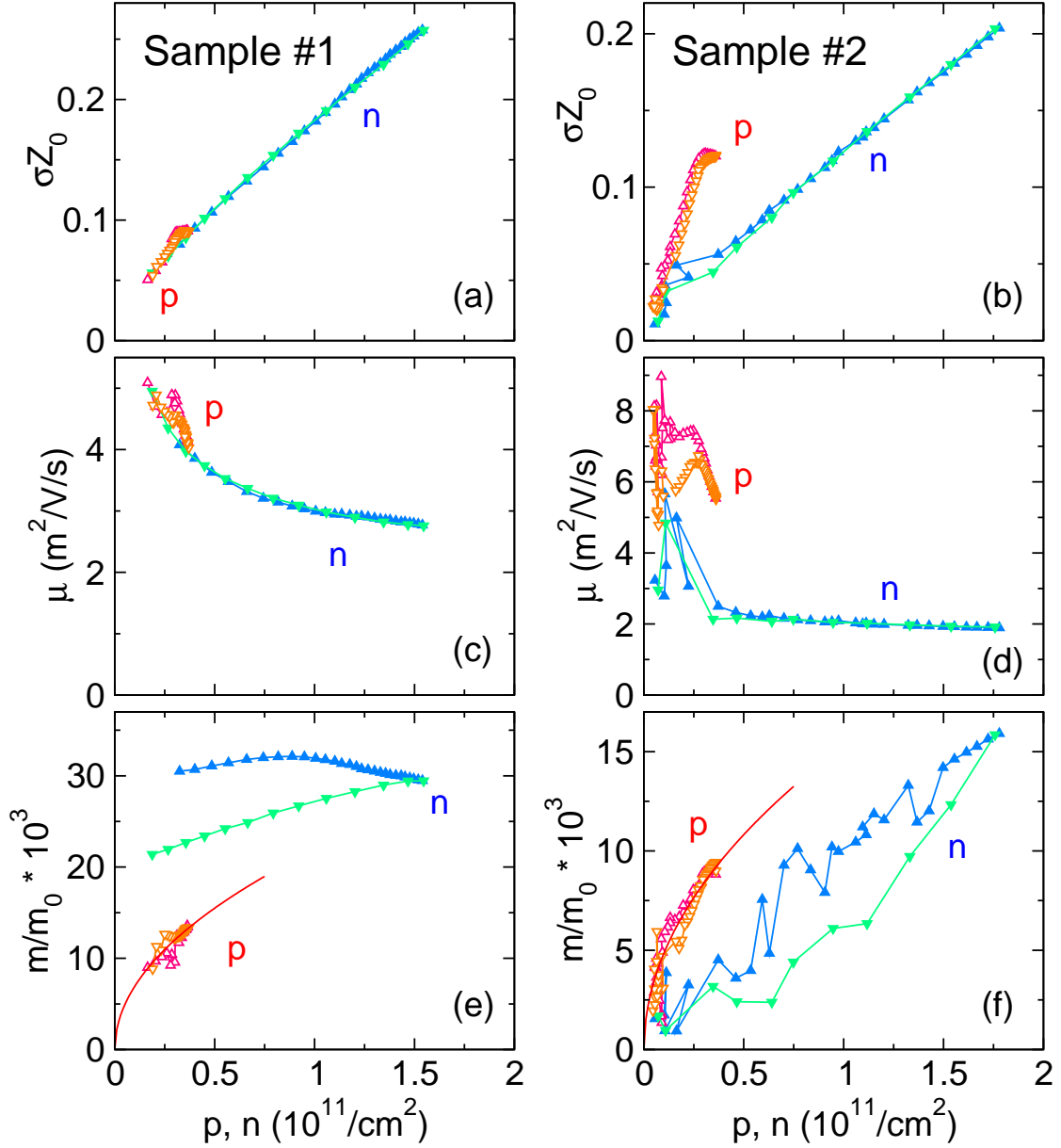}
\caption{\textbf{Electrodynamic parameters of the charge carriers}. The data are obtained from simultaneous fitting of the magneto-optical spectra using the Drude model in the presence of the external magnetic field~\cite{shuvaev_sst_2012, dziom_2d_2017}. Left panels - sample \#~1 (340 GHz), $d=6.3$~nm, right panels - sample \#~2 (320 GHz), $d=6.6$~nm. (a,b) - normalized conductivity, (c,d) - mobility, (e,f) - effective mass. Up triangles - increasing gate voltage, down triangles - decreasing voltage, closed green/blue triangles - electrons, open red/orange triangles - holes.  Blue and green lines are to guide the eye, red solid lines in (e,f) represent the square root behavior expected for Dirac dispersion.} \label{figpar}
\end{figure}

From fitting the experimental transmission data, the parameters of the charge carriers are obtained and they are shown in Fig.~\ref{figpar}. As expected, the static conductivity (panels (a,b)) increases approximately linearly with the density of electrons and holes, indicating a rough independence of the mobility on the carriers concentration. The effective cyclotron mass demonstrates generally complicated behavior, reflecting non-parabolic band structure of our HgTe samples. Solely, the approximate square root-like curvature for the hole masses of the sample \#2 is a clear indication of the Dirac dispersion of these type of carriers, similar to graphene \cite{novoselov_nature_2005, neto_rmp_2009} and to HgTe with critical thickness~\cite{dziom_2d_2017, olbrich_prb_2013, ikonnikov_sst_2011, zoth_prb_2014, ludwig_prb_2014, zholudev_prb_2012}. In addition, we note the enhanced hole mobility in Fig.~\ref{figpar}(d). In agreement with the magnetoresistance experiments this effect is due to screening of the fluctuation potential by hole valley reservoirs~\cite{kozlov_jetpl_2013, kozlov_jetpl_2015}.

\begin{figure}[tbp]
\includegraphics[width=1.0\linewidth]{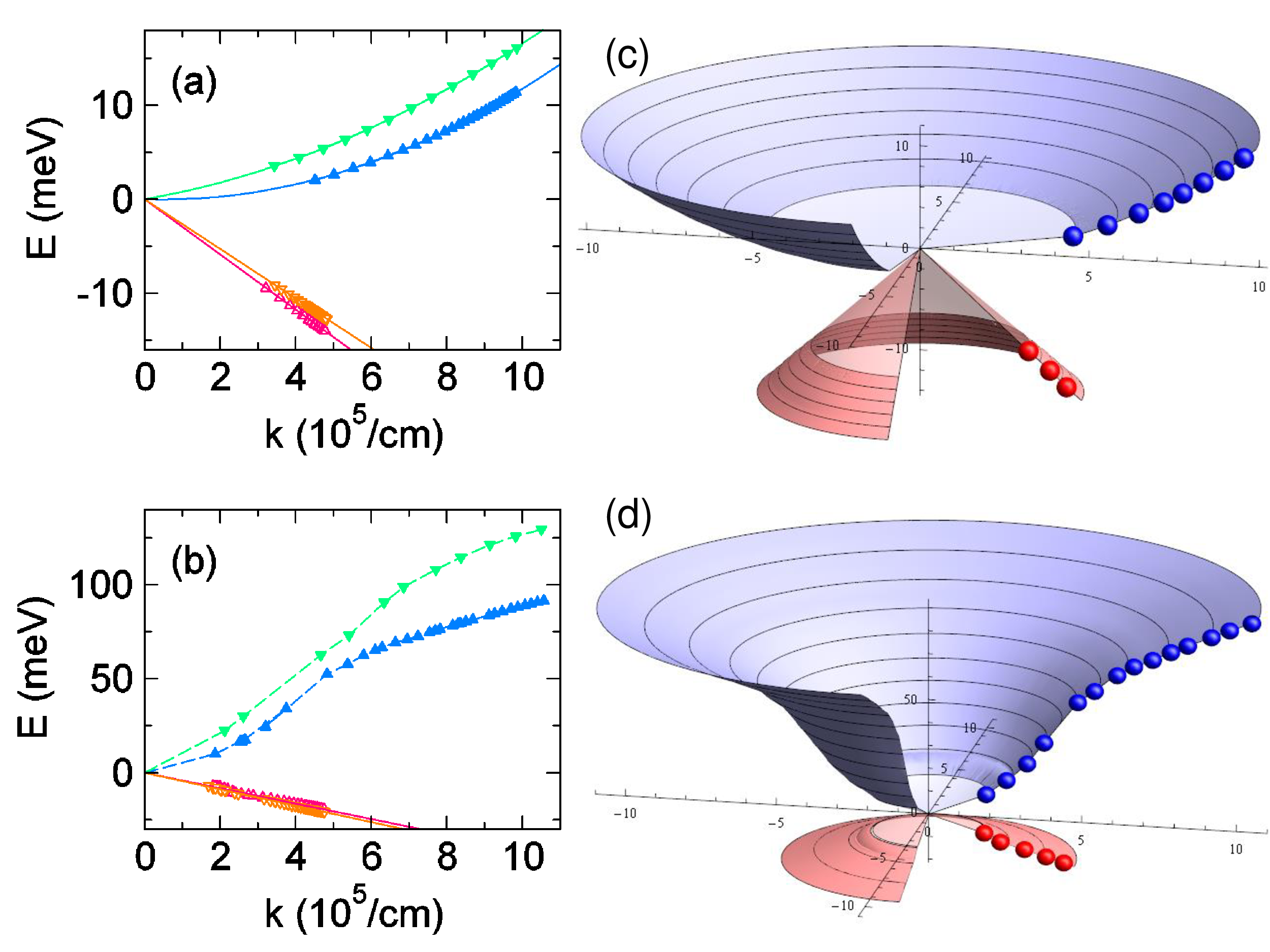}
\caption{\textbf{Band structure of the HgTe films with critical thickness.} (a,c) - Sample \#1, (b,d) - sample \#2. Left panels - 2D presentation. The notations are the same as in Fig.~\ref{figpar}. Right panels - the data is replotted in a 3D form.} \label{figband}
\end{figure}

\subsection{Band structure}

HgTe quantum wells with critical thickness are on the border between normal and inverted dispersion relations. As a result, they may be characterized by a zero-gap band structure with linear momentum-energy relations (Dirac cone)~\cite{bernevig_science_2006, qi_prb_2008, buttner_nphys_2011}. Previously, several experimental techniques have been utilized to confirm this band structure and to investigate the deviations from linear dispersion. Measurements of cyclotron resonance in these systems are especially helpful, as they are directly connected with the band structure via Eq.~(\ref{eqCR2}).

For linear dispersion relations, $E=\hbar v_f k$, Eq.~(\ref{eqCR2}) predicts~\cite{neto_rmp_2009, dziom_2d_2017} that the cyclotron mass is proportional to the square root of the 2D charge density, which was confirmed experimentally~\cite{ikonnikov_sst_2011, kvon_jetpl_2012, olbrich_prb_2013, zoth_prb_2014, dziom_2d_2017}. For HgTe with thickness away from critical, additional quadratic terms in the dispersion are necessary to explain the cyclotron mass~\cite{kvon_physe_2008, ikonnikov_sst_2011, zoth_prb_2014}. The results by cyclotron resonance at low frequencies are in line with the magnetospectroscopy at higher magnetic fields~\cite{zholudev_nrl_2012, zholudev_prb_2012, ludwig_prb_2014, ikonnikov_prb_2016}. These experiments are done in quantum regime and the transitions between separate Landau levels can be investigated making the effects of the width fluctuations detectable~\cite{zholudev_nrl_2012}. In additions, external magnetic fields above $B \gtrsim 1$~T may substantially shift the electronic structure, e.g. leading to transition from normal to inverted bands~\cite{ikonnikov_prb_2016}. We note that in present experiments the cyclotron resonance appear below $B \lesssim 0.5$~T. Therefore, in a first approximation the distortions by magnetic fields may be neglected.

Besides cyclotron experiments, the effective mass can be obtained via the Shubnikov--de Haas (SdH) effect~\cite{novoselov_nature_2005, minkov_prb_2013, minkov_prb_2014, minkov_prb_2015, minkov_prb_2016}. From the fitting of SdH results the electronic energy spectrum can be modelled. In addition, for HgTe with thickness close to critical, a spin-orbit splitting has been detected for hole bands~\cite{minkov_prb_2016}. In present experiments we do not see such splitting neither in low fields nor in the quantum regime~\cite{shuvaev_prl_2016}, which might be due to the weakness of the cyclotron signal from the split bands.

Recently a capacitance spectroscopy provided useful insights into the band structure of HgTe quantum wells~\cite{kozlov_jetpl_2016, kozlov_prl_2016}. In these experiments a direct access to the density of states is obtained via precise measurements of the sample capacitance as a function of the magnetic field and the gate voltage. In case of the HgTe wells with critical thickness~\cite{kozlov_jetpl_2016} a Dirac dispersion on the electronic part of the spectrum has been observed. At low energies $E \lesssim 30$~meV the deviations from the linear dispersion have been detected and attributed to disorder effects.

The concentration dependence of the cyclotron masses in the present samples (Fig.~\ref{figpar}) allows an insight into the band structure as shown in Fig.~\ref{figband}. The 2D presentations in the left panels are obtained by directly integrating Eq.~(\ref{eqCR2}). We recall here that the isotropy within the 2D plane is assumed. Within the same approximations, the band structure is rotation-symmetric and, therefore, the 3D surface plot may be generated as shown in the right panels.

For two samples in Fig.~\ref{figband}, the dispersion of the hole carriers show nearly perfect linear behavior at low energies. In present experiments we probably do not reach the heavy hole bands due to large density of states at higher hole energies. Therefore, only linear part of the spectrum remain visible. In addition, in this regime the holes are highly mobile due to the mentioned hole valley reservoirs~\cite{kozlov_jetpl_2013, kozlov_jetpl_2015} which screen the fluctuation potential.

The inspection of the data in Fig.~\ref{figband} reveals substantial quadratic curvature of the electronic dispersion, as demonstrated by the solid lines in Fig.~\ref{figband}(a) for sample \#1. This agrees with the fact that with $d=6.3$\,nm, this sample is slightly thinner than the critical thickness $d_c=6.6$\,nm for films on GaAs substrates~\cite{ludwig_prb_2014}. As discussed above, deviations from critical thickness~\cite{ikonnikov_sst_2011, zoth_prb_2014} and thickness fluctuations~\cite{kozlov_jetpl_2016} are the most probable reasons to the observed parabolic corrections. Assuming that the thickness fluctuations are of the order of the lattice constant (0.65\,nm), they would play an important role for film thickness close to critical. It may be expected that the effect of fluctuations gradually disappears for thicknesses above 20\,nm. In that case, 2D samples with linear dispersion still may be available using Cd-doped HgTe.

\begin{figure}[tbp]
\includegraphics[width=.99\linewidth]{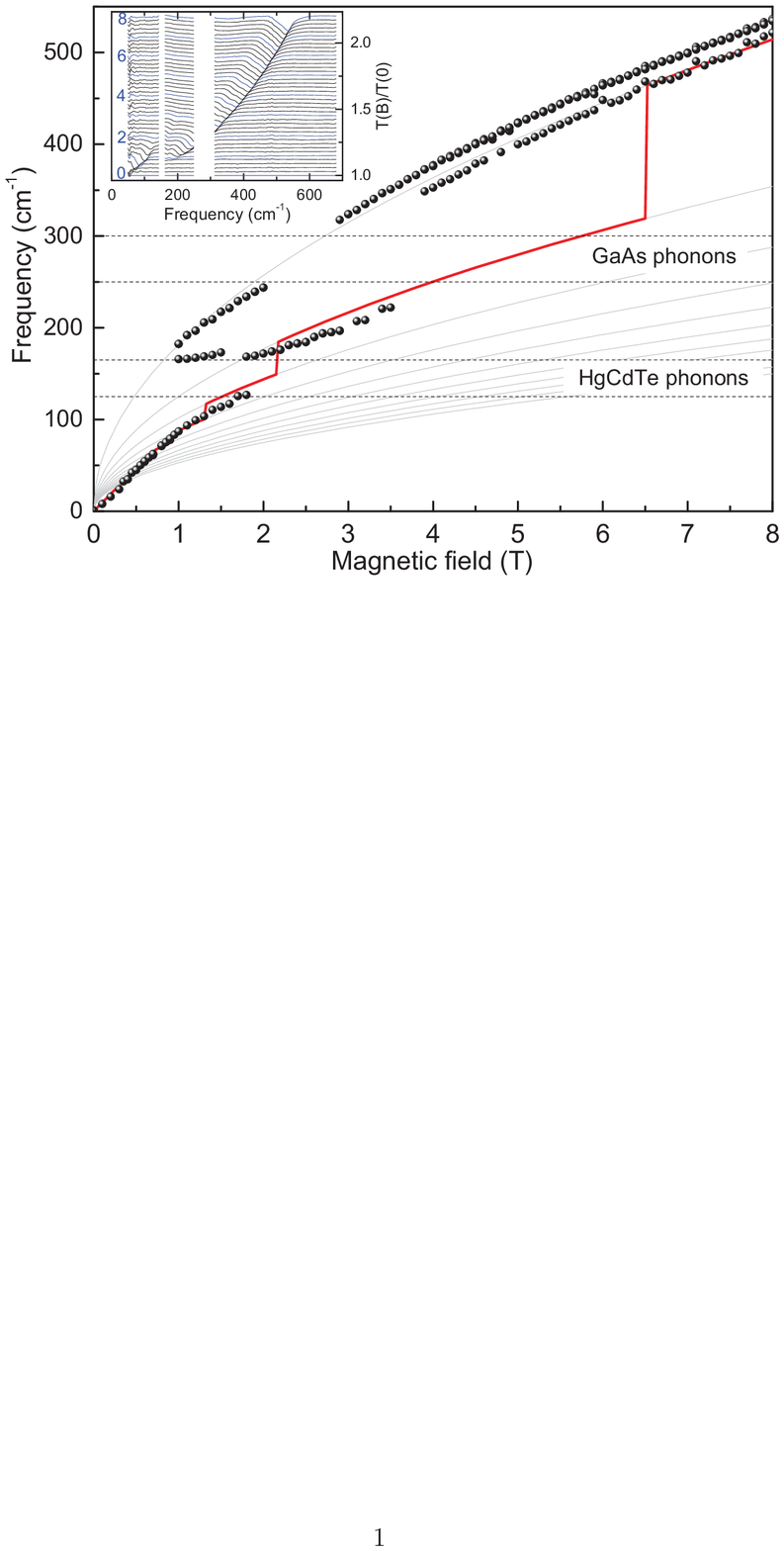}
\caption{\textbf{Far-infrared magneto-optical spectra.} Symbols - experimental absorption frequencies in the infrared transmission experiment. Solid gray lines - magnetic field dependence of the Landau level energy. Red solid line - calculated position of the absorption frequencies as described in the text. The inset shows examples of the normalized transmission spectra. The curves are shifted for clarity.} \label{figir}
\end{figure}

\subsection{Infrared magneto-optics}

The terahertz analysis described above has been extended to the infrared frequency range~\cite{shao_nanol_2017}. These experiments have been performed on sample \#3 with $d=6.6$~nm, which is closely similar to sample \#2. The carrier density was constant in these experiments and it was estimated from the transmission intensity as $n=7\cdot 10^{10}$~cm$^{-2}$. As demonstrated in Fig.~\ref{figir}, in high magnetic fields $B>1$~T, the classical cyclotron resonance mode is split into transitions between single Landau levels. In addition, for magnetic fields above $B\sim 4$~T, two modes with close-by frequencies are observed, which indicates that the initial twofold degeneracy of the Dirac cones is lifted in the high magnetic fields~\cite{ludwig_prb_2014}. Characteristic gaps in the data close to \cm{150}~ and \cm{270} are due to phonon absorbtion in GaAs~\cite{hass_jpcm_1962} substrate and HgCdTe buffer layers~\cite{sheregii_prl_2009} which absorb the infrared radiation. In addition, because the HgTe film and the HgCdTe layers are close to each other, the electron gas do interact with HgCdTe phonons around  \cm{150}. This interaction is seen as an effect of avoided crossing of cyclotron resonance frequency close to \cm{150}.

The gray solid lines in Fig.~\ref{figir} correspond to the field-dependence of the transition energies between Landau levels~\cite{mcclure_pr_1956} with $E_n=\mathrm{sign}(n)v_F\sqrt{2e\hbar B |n|}$ in the simple model of an undistorted Dirac cone. Fig.~\ref{figir} demonstrates that the HgTe film with critical thickness may indeed be reasonably described within this approximation. The red solid line demonstrates the field positions of the ${n\rightarrow n+1}$ transitions at the Fermi energy. From the fits to the experimental data, we determine the Fermi velocity as $v_F = 1.2\cdot 10^6 cm/s$. This agrees well with that obtained from the slope of dispersion relationship in the similar sample \#2 in Fig.~\ref{figband}(b), $v_F = 1.1\cdot 10^6 cm/s$. In the present case the absolute values are about 40\% higher than in other cyclotron resonance~\cite{olbrich_prb_2013, ludwig_prb_2014} and in magnetocapacitance~\cite{kozlov_jetpl_2016} experiments. This may be attributed to parabolic corrections of the band structure which lead to a concentration dependence of the Fermi velocity.

The existence of parabolic corrections agrees with the dispersion relation for a similar sample \#2 in Fig.~\ref{figband}(b). The estimated electron density for sample \#3, $n=7\cdot 10^{10}$~cm$^{-2}$, would  correspond to $k_F \approx 6.6 \cdot 10^{5}$\cm{}, i.e., the point with visible nonlinear curvature. Another indication of deviations from ideal Dirac dispersion is the splitting of the absorption mode in high fields, Fig.~\ref{figir}. Such splitting has been recently observed in magneto-optical experiments on similar samples~\cite{ludwig_prb_2014}, and explained on the basis of an effective Dirac model with nonzero gap. However, we note that simple Dirac picture still provide a reasonable description of the data in Fig.~\ref{figir}.

\section{Conclusions}

In conclusion, we directly obtained the band structure of a two-dimensional Dirac semi-metal from the doping dependence of the cyclotron resonance. We observe a linear Dirac-like dispersion on the hole side of the band structure and detectable quadratic corrections for the electrons. This procedure to obtain the band structure is especially useful for thin films where protective layers impede such standard techniques as angular resolved photoemission spectroscopy (ARPES).

\subsubsection*{Acknowledgments}

We acknowledge valuable discussion with G. Tkachov. This
work was supported by Austrian Science Funds (W-1243, P27098-N27), by DOE-BES (DE-FG02-00ER45799), and by the Moore Foundation (Grant GBMF4533)

\bibliography{literature}

\end{document}